\documentclass[preprint2,twocolumn,times,tighten]{aastex631}
\usepackage{subfigure}

\usepackage{amsmath}
\usepackage{hyperref}
\usepackage{amssymb}
\usepackage{natbib}
\usepackage{morefloats}
\usepackage{multirow}
\usepackage{array}
\usepackage{verbatim}
\usepackage{color}

\begin{document}

\title{Probing Three-Dimensional Magnetic Fields: III - Synchrotron Emission and Machine Learning}

\email{yuehu@ias.edu; alazarian@facstaff.wisc.edu; *NASA Hubble Fellow}

\author[0000-0002-8455-0805]{Yue Hu*}
\affiliation{Institute for Advanced Study, 1 Einstein Drive, Princeton, NJ 08540, USA }
\author{A. Lazarian}
\affiliation{Department of Astronomy, University of Wisconsin-Madison, Madison, WI 53706, USA}


\begin{abstract}
Synchrotron observation serves as a tool for studying magnetic fields in the interstellar medium and intracluster medium, yet its ability to unveil three-dimensional (3D) magnetic fields, meaning probing the field'splane-of-the-sky (POS) orientation, inclination angle relative to the line of sight, and magnetization from one observational data, remains largely underexplored. Inspired by the latest insights into anisotropic magnetohydrodynamic (MHD) turbulence, we found that synchrotron emission's intensity structures inherently reflect this anisotropy, providing crucial information to aid in 3D magnetic field studies: (i) the structure's elongation gives the magnetic field's POS orientation and (ii) the structure's anisotropy degree and topology reveal the inclination angle and magnetization. Capitalizing on this foundation, we integrate a machine learning approach-Convolutional Neural Network (CNN)-to extract this latent information, thereby facilitating the exploration of 3D magnetic fields. The model is trained on synthetic synchrotron emission maps, derived from 3D MHD turbulence simulations encompassing a range of sub-Alfv\'enic to super-Alfv\'enic conditions. We show that the CNN is physically interpretable and the CNN is capable of obtaining the POS orientation, inclination angle, and magnetization. Additionally, we test the CNN against the noise effect and the missing low-spatial frequency. We show that this CNN-based approach maintains a high degree of robustness even when only high-spatial frequencies are maintained. This renders the method particularly suitable for application to interferometric data lacking single-dish measurements. We applied this trained CNN to the synchrotron observations of a diffuse region. The CNN-predicted POS magnetic field orientation shows a statistical agreement with that derived from synchrotron polarization.
\end{abstract}

\keywords{Interstellar magnetic fields (845) --- Interstellar synchrotron emission(856) --- Magnetohydrodynamics (1964) --- Convolutional neural networks(1938)}


\section{Introduction}

Synchrotron radiation, emanating from relativistic electrons gyrating around magnetic field lines \citep{1979rpa..book.....R,1992ARA&A..30..575C}, is a probe of magnetic fields in interstellar medium (ISM) and intracluster medium (ICM) \citep{1965ARA&A...3..297G,2008A&A...477..573S,2016A&A...594A..10P,2019Sci...364..981G,2020ApJ...890...70W,2022ApJ...925..165H,2020ApJ...901..162H,2024NatCo..15.1006H}. This radiation not only facilitates the estimation of magnetic field strengths at equipartition \citep{1994ApJ...421..225C,2009A&A...494...21A,2022ApJ...925L..18Y,2024MNRAS.527.1275Y}, which is pivotal for elucidating cosmic ray acceleration mechanisms \citep{1966ApJ...146..480J,1978MNRAS.182..443B,2012SSRv..166...71B,2014ApJ...783...91C,2014ApJ...785....1B,2022ApJ...925...48X,2022ApJ...934..136X}, but it also allows the determination of magnetic field orientations \citep{2001SSRv...99..243B,2015A&ARv..24....4B,2016A&A...594A..10P,2019MNRAS.486.4813Z,2021ApJ...920....6G}. This is crucial for understanding different physics across scales, from large-scale galaxy clusters \citep{2004IJMPD..13.1549G,2014IJMPD..2330007B,2021MNRAS.502.2518S,2024NatCo..15.1006H}, galaxies \citep{2001SSRv...99..243B,2015A&ARv..24....4B,2023ApJ...946....8T}, to small- scale individual supernova remnants \citep{1983Natur.304..243M,2008A&A...482..783X,2009A&A...503..827X,2012SSRv..166..231R}. Despite its critical role, our comprehension of synchrotron radiation and the magnetic field insights it can offer is still evolving.

Extracting three-dimensional (3D) magnetic field information from synchrotron radiation poses a substantial challenge. Polarized synchrotron emission offers two-dimensional (2D) insights into the magnetic field orientation within the plane of the sky (POS) \citep{1979rpa..book.....R,1983Natur.304..243M,2001SSRv...99..243B,2009A&A...503..827X,2012SSRv..166..231R,2015A&ARv..24....4B,2016A&A...594A..10P,2019MNRAS.486.4813Z,2021ApJ...920....6G}, but it cannot directly probe the POS direction and the inclination angle of the magnetic field relative to the line of sight (LOS). The scenario is further complicated by the Faraday rotation effect, which alters the intrinsic polarization angle of the emission sources \citep{2007ASPC..365..242H,2009ApJ...702.1230T,2012A&A...542A..93O,2016ApJ...824..113X,2019A&A...632A..68T}. Consequently, not only is accurately measuring the POS magnetic field from synchrotron polarization challenging, but tracing the actual 3D magnetic field structure becomes a formidable task.

Recent advancements have unlocked the potential of using anisotropy in synchrotron radiation to trace 3D magnetic fields (i.e., determining the POS orientation, inclination angle, and magnetization simultaneously). The theory that relates the anisotropy of synchrotron radiation with the properties of magnetohydrodynamic (MHD) turbulence \citep{GS95,LV99} was formulated in \citet{2012ApJ...747....5L}. The anisotropy means the observed synchrotron intensity structures tend to elongate along the magnetic field lines intersecting them. The elongation therefore can be used as a probe of the magnetic field orientation. Based on this property, \citet{2017ApJ...842...30L} introduced the Synchrotron Intensity Gradients (SIG) to trace the POS magnetic field orientation \citep{2017ApJ...842...30L,2024NatCo..15.1006H}. The technique was demonstrated to be applicable (with an uncertainty less than 8$^\circ$) for application under both sub-Alfv\'enic and super-Alfv\'enic conditions, notably within galaxy clusters as demonstrated by \citet{2024NatCo..15.1006H}. Crucially, subsequent research by \citet{HLX21a} and \citet{2024MNRAS.52711240H} has shown that the observed anisotropy in the POS contains information on the underlying 3D magnetic field structures. This insight stems from the fact that the anisotropy, or elongation along the magnetic field line, is inherently a 3D phenomenon. Therefore, the observed POS anisotropy of synchrotron intensity structures is influenced by the projection effect, which is determined by the field's inclination angle, and the magnetization \footnote{Magnetization is defined as $1/M_A$, where $M_A$ is the Alfv\'en Mach number. Large $M_A$ suggest a relatively weak magnetic field, thereby a weak magnetization.} level of the medium.

Given these theoretical considerations, the observed structure of synchrotron emission intrinsically encompasses information pertinent to 3D magnetic fields. In this study, we propose the employment of a machine learning paradigm—specifically, Convolutional Neural Networks (CNNs; \citealt{lecun1998gradient})—to extract spatial features within the synchrotron intensity maps, thereby facilitating the measurement of 3D magnetic fields. This includes determining the orientation of the magnetic field within the POS, ascertaining the magnetic field's inclination angle, and assessing the overall magnetization. A similar method employing CNNs for 3D magnetic field tracing has been previously proposed by \citet{2024MNRAS.52711240H}, affirming the CNN's capability to identify magnetic-field-specific spatial features from spectroscopic data, thereby yielding precise measurements in star-forming regions. The physical conditions for these regions correspond to supersonic cold gas, which contrasts subsonic warm/hot gas in typical synchrotron-emitting regions. This work aims to extend the CNN method to trace the 3D magnetic field in diffuse synchrotron-emitting regions, opening a new way of using vast data sets of diffuse synchrotron emission to get information unavailable through a traditional synchrotron data analysis.

Crucially, our approach transcends mere algorithmic application; we aim to understand which synchrotron intensity features are indicative of magnetic field properties, why these features are significant, and the fundamental physical principles they represent. This strategy not only deepens our understanding of CNN's efficacy in producing 3D magnetic field mappings but also the relation between observed synchrotron structures and magnetic field properties. For the CNN training, we utilize 3D MHD subsonic simulations that encompass a range of magnetization levels, ranging from sub-Alfv\'enic conditions (i.e., strong magnetic fields), through trans-Alfv\'enic, to super-Alfv\'enic scenarios (i.e., weak magnetic fields). These simulations are subsequently post-processed to create synthetic synchrotron observations. The use of synthetic observations in our study is crucial due to the inherent limitations of current observational data in providing 3D magnetic field information. Synthetic observations allow us to validate our technique under controlled conditions where the magnetic field structure is known prior.

This paper is organized as follows. In \S~\ref{sec:theory}, we outline the fundamental aspects of MHD turbulence anisotropy observed in synchrotron emissions and their association with 3D magnetic field orientation and overall magnetization. \S~\ref{sec:data} provides a description of the 3D MHD simulations and the synthetic observations utilized in this study, alongside details of our CNN model. In \S~\ref{sec:result}, we present the results of our numerical testing and observational application. \S~\ref{sec:dis} delves into discussions surrounding the uncertainties and future prospects of employing machine learning techniques for astrophysical analysis.
We conclude with a summary of our findings in \S~\ref{sec:con}.

\section{Theoretical consideration}
\label{sec:theory}
\subsection{Anisotropy in MHD turbulence}
A significant advancement in our understanding of MHD turbulence was the introduction of the "critical balance" condition, equating the cascading time, $(k_\bot \delta v_l)^{-1}$, with the wave periods, $(k_\parallel v_A)^{-1}$, as proposed by \citet{GS95} (hereafter GS95). Here, $k_\parallel$ and $k_\bot$ denote the components of the wavevector parallel and perpendicular to the magnetic field, respectively. The term $\delta v_l$ refers to the turbulent velocity at scale $l$, and $v_A = B/\sqrt{4\pi\rho}$ represents the Alfv\'en speed, where $B$ is the magnetic field strength and $\rho$ is the gas mass density. It is essential to note that GS95's analysis is grounded in a global reference frame, wherein the orientation of wavevectors is defined relative to the mean magnetic field.

\citet{LV99} (hereafter LV99) subsequently demonstrated that the "critical balance" condition also holds in a local reference frame, defined relative to the magnetic field intersecting an eddy at scale $l$. According to LV99, the process of turbulent reconnection of the magnetic field \footnote{Note that reconnection happens not just at a specific scale and phase, but everywhere the magnetic field lines get intersected. Turbulence can induce fluctuations in field lines, so reconnection happens.}, occurring within one eddy turnover time, facilitates the mixing of magnetic field lines perpendicular to the magnetic field's orientation. This mixing induces changes in fluid velocities perpendicular to the magnetic field lines, ensuring that the motion of eddies sized $l_\bot$ and oriented perpendicular to the local magnetic field direction is not suppressed. This implies that the perpendicular direction poses minimal resistance to turbulent cascading, which typically follows the Kolmogorov law.
\begin{figure*}
	\includegraphics[width=1.0\linewidth]{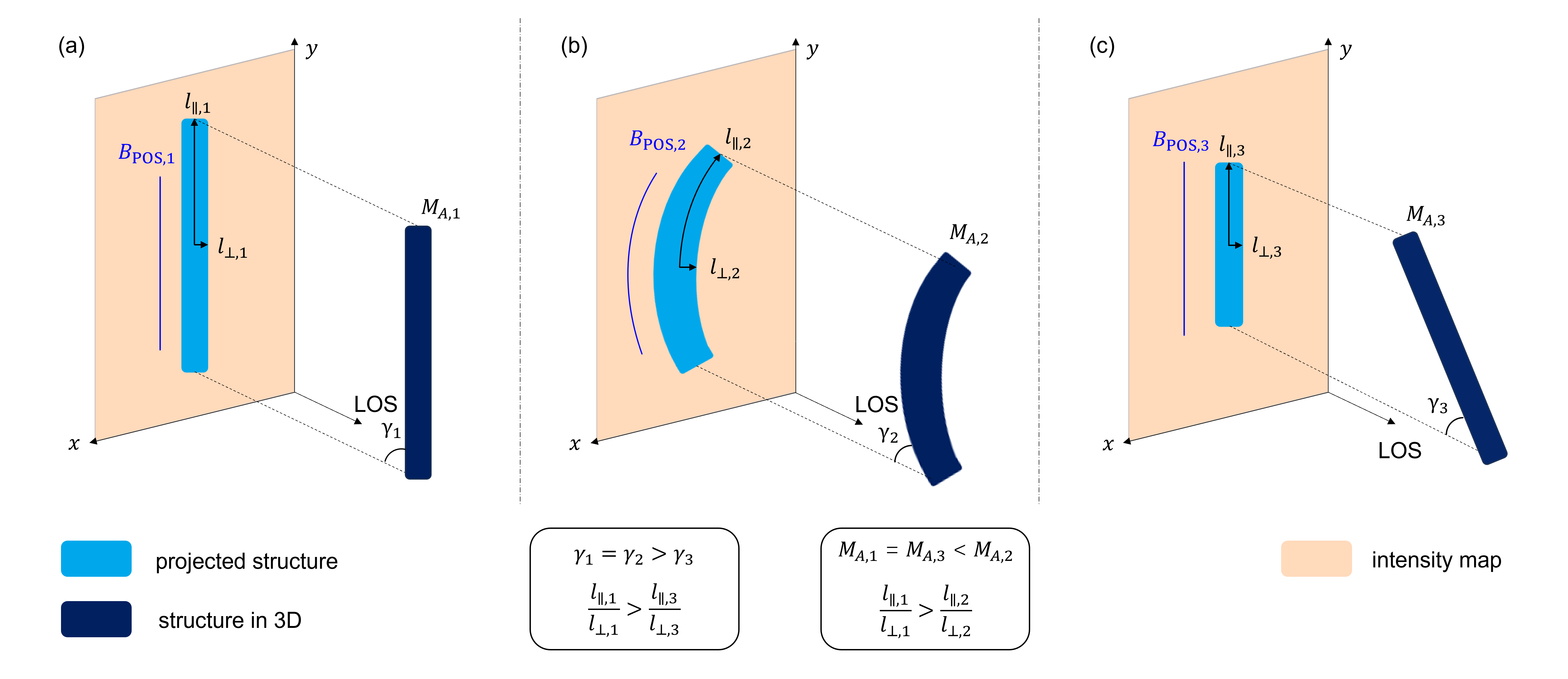}
    \caption{An Illustration of how the observed synchrotron intensity structures are regulated by the Alfv\'en Mach number $M_A$ and inclination angle $\gamma$. Within all three panels, these intensity structures elongate along the POS magnetic field where $l_\parallel>l_\bot$. Structures 1 and 2, depicted in panels (a) and (b), are projected onto the POS with identical inclination angles $\gamma_1=\gamma_2$, yet exhibit different magnetization with $M_{A,1}^{-1}>M_{A,2}^{-1}$. Notably, the anisotropy observed, represented as $l_\parallel/l_\bot$, in the weakly magnetized Structure 2 is less pronounced than in Structure 1. Structure 2 is less straightened because the weak magnetic field has more fluctuations. Comparatively, Structures 1 and 3—showcased in panels (a) and (c)—possess equivalent magnetization $M_{A,1}^{-1}=M_{A,3}^{-1}$, but divergent inclination angles with $\gamma_1>\gamma_3$. The observed anisotropy decreases with smaller  $\gamma$, though it is crucial to note that the straightness of Structure 3 remains unaffected by this projection. Modified from \citet{2024MNRAS.52711240H}.}
    \label{fig:anisotropy}
\end{figure*}

Considering the "critical balance" condition in the local reference frame: 
$\delta v_{l,\bot}l_\bot^{-1}\sim v_Al_\parallel^{-1}$ and the Kolmogorov relation in strong turbulence regime (i.e., $\delta v_{l,\bot} = (\frac{l_\bot}{L_{\rm inj}})^{1/3}\delta v_{\rm inj}M_A^{1/3}$, where $\delta v_{\rm inj}$ is the injection velocity at injection scale $L_{\rm inj}$ and $\delta v_{l,\bot}$ is turbulent velocity along the direction perpendicular to the magnetic field at scale $l$), one can get the scale-dependent anisotropy scaling \citep{LV99}:
\begin{equation}
	\begin{aligned}
	 l_\parallel= L_{\rm inj}(\frac{l_\bot}{L_{\rm inj}})^{\frac{2}{3}} M_A^{-4/3},~~~M_A\le 1,\\
	\delta v_{l,\bot}= \delta v_{\rm inj}(\frac{l_\bot}{L_{\rm inj}})^{\frac{1}{3}}M_A^{1/3}, ~~~M_A\le 1,
	\label{eq.lv99}
	\end{aligned}
\end{equation}
where $l_\bot$ and $l_\parallel$ represent the perpendicular and parallel scales of eddies with respect to the local magnetic field, respectively. $M_A=\delta v_{\rm inj}/v_A$ is the Alfv\'en Mach number. This scaling relation has been demonstrated by numerical simulations \citep{2000ApJ...539..273C, 2001ApJ...554.1175M, 2003MNRAS.345..325C, 2010ApJ...720..742K, HXL21,2022ApJ...941..133H,2024MNRAS.527.3945H} and in-situ measurements in the solar wind \citep{2016ApJ...816...15W, 2020FrASS...7...83M, 2021ApJ...915L...8D, 2023arXiv230512507Z}.

Eq.~\ref{eq.lv99} provides the scaling relation for velocity fluctuations and reveals the anisotropic nature of turbulent eddies (i.e., $l_\parallel \gg l_\bot$). In other words, the perpendicular velocity fluctuation is more significant than the parallel fluctuations at the same scale \citep{HXL21}. The relationships for density and magnetic field fluctuations can also be derived using the linearized continuity and induction equations, considering the components as a sum of their mean and fluctuating parts: 
$\rho=\rho_0+\delta\rho_l$, $\pmb{v}=\pmb{v}_0+\delta \pmb{v}_l$, and $\pmb{B}=\pmb{B}_0+\delta \pmb{B}_l$, where $\rho_0$ and $\pmb{B}_0$ denote the mean density and mean magnetic field strength, while the mean velocity field $\pmb{v}_0=0$ \citep{2003MNRAS.345..325C}. The equations in Fourier space are:
\begin{equation}
	\begin{aligned}
        \omega \delta\rho_k = \rho_0 \pmb{k}\cdot \delta\pmb{v}_k,\\
        \omega\delta \pmb{B}_k = \pmb{k}\times(\pmb{B}_0\times \delta\pmb{v}_k),
	\end{aligned}
\end{equation}
in which dispersion relation for Alfv\'enic turbulence is $\omega/k=v_A$.  Considering the displacement vector $\pmb{\xi}$, where the time derivative of $\pmb{\xi}$ gives the velocity vector $\partial\pmb{\xi}/\partial t = \pmb{v}=v\hat{\pmb{\xi}}$, we obtain:
\begin{equation}
	\begin{aligned}
		\delta \rho_l&= \delta v_l\frac{\rho_0}{v_A}\mathcal{F}^{-1}(|\hat{\pmb{k}}\cdot\hat{\pmb{\xi}}|),\\
		\delta B_l&= \delta v_l\frac{B_0}{v_A}\mathcal{F}^{-1}(|\hat{\pmb{B}}_0\times\hat{\pmb{\xi}}|), 
		\label{eq.rhoB}
	\end{aligned}
\end{equation}
 where $\hat{\pmb{k}}$ and $\hat{\pmb{\xi}}$ represent the unit wavevector and displacement vector, respectively. $\mathcal{F}^{-1}$ denotes the inverse Fourier transform. The density and magnetic field fluctuations induced by turbulence are proportional to the velocity fluctuations, and dominated by their perpendicular components.
 
In addition to the anisotropy, the topology of magnetic field lines is regulated by the magnetization. Within a domain of strong magnetization, magnetic field lines exhibit minimal deviation, attributed to subdued fluctuations, resulting in predominantly straightened topology. In contrast, a weaker magnetic field, signified by a higher $M_A$, is associated with stronger orientation fluctuations in the magnetic field. This enhancement leads to field lines adopting more curved formations \citep{2020ApJ...898...66Y}. Together with Eq.~\ref{eq.lv99}, we have three important properties of MHD turbulence \citep{2024MNRAS.52711240H}: 
\begin{enumerate}
    \item Turbulent eddies predominantly stretch along the local magnetic field (i.e., $l_\parallel \gg l_\bot$), underscoring an anisotropy in velocity, density, and magnetic field structures.
    \item The degree of anisotropy, quantified as $l_\parallel/l_\bot$, is intricately linked to the magnetization, represented by $M_A^{-1}$.
    \item Difference in magnetization manifests distinctively in the magnetic field topology.
\end{enumerate}

Note that for super-Alfv\'enic scenarios where $M_A \gg 1$, turbulence approaches isotropy, dominated by hydrodynamic turbulence. However, turbulence cascades energy from larger injection scales down to smaller scales and progressively diminishes turbulent velocity. 
Assuming Kolmogorov turbulence, the magnetic field's energy approaches that of turbulence (i.e., the Alfv\'en Mach number becomes unity) at the transition scale $l_A$, which can be derived from (see \citealt{2006ApJ...645L..25L}):
\begin{equation}
\begin{aligned}
     \frac{1}{2}\rho(\frac{l_A}{L_{\rm inj}})^{2/3}\delta v_{\rm inj}^2&=\frac{1}{8\pi}B^2,\\
    l_A&= L_{\rm inj}/M_A^3,
\end{aligned}    
\end{equation}
below which the magnetic field's role becomes important and the anisotropy can be observed (see Eq.~\ref{eq.lv99}).

\subsection{Anisotropy in synchrotron emission}
The intrinsic relationship between synchrotron emission and the density of relativistic electrons and magnetic fields (see Eq.~\ref{eq.iqu}), ensures that the anisotropy and magnetic field topology are naturally encoded in the observed synchrotron intensity structure. The observed intensity $I(x,y)$ are expressed as \citep{1986rpa..book.....R,1970ranp.book.....P,2016ApJ...831...77L}:
\begin{equation}
\label{eq.ip}
I(x,y) \propto \int (j_\bot + j_\parallel) dz,
\end{equation}
where $j_\bot$ and $j_\parallel$ denote synchrotron emissivities perpendicular and parallel to the POS magnetic field, respectively. Further expansion of $j_\bot$ and $j_\parallel$ reveals the intrinsic synchrotron emission $I_i(x,y,z) = j_\bot+j_\parallel$, as follows:
\begin{equation}
\label{eq.iqu_i}
I_i(x,y,z) \propto n_{e} (B_x^2 + B_y^2)B_{\perp}^\alpha ,
\end{equation}
where $B_{\perp}=\sqrt{B_x^2 + B_y^2}$ represents the magnetic field component perpendicular to the LOS, with $B_x$ and $B_y$ as its $x$ and $y$ components, respectively. $n_e$ indicates the relativistic electron number density and $\alpha$ denotes synchrotron emission index.

As indicated in Eq.~\ref{eq.rhoB}, when describing the density and magnetic field as a sum of their mean and fluctuating components, their fluctuations are predominantly perpendicular to the magnetic field. The expressions in Eq.~\ref{eq.iqu_i} suggest that fluctuations in synchrotron emission intensity are determined by these in the magnetic field and density. Other constant factors are not explicitly detailed in Eq.~\ref{eq.iqu_i}, as they do not alter the characteristics of these fluctuations. Consequently, the fluctuations in $I_i$ exhibit pronounced anisotropy, showing more significant fluctuations perpendicular to the magnetic field.

This anisotropy implies that the contours of synchrotron intensity,
as illustrated in Fig.~\ref{fig:anisotropy}, are elongating along the magnetic fields. The elongation in the projected intensity structure unveils the orientation of the POS magnetic field, while more information is needed to infer the POS direction.

\textbf{Projection effect:} The observed synchrotron intensity is subject to the projection along the LOS. The projection effect changes the observed anisotropy degree, defined as $l_\parallel/l_\bot$. As shown in Fig.~\ref{fig:anisotropy}, a large inclination angle of the magnetic field indicates a weak projection effect, resulting in a larger $l_\parallel/l_\bot$ ratio. Conversely, a small inclination angle reduces the parallel scale $l_\parallel$, thus diminishing the anisotropy degree. Therefore, the observed anisotropy degree provides insights into the magnetic field’s inclination angle.

Additionally, fluctuations cause the magnetic field lines to exhibit curvature, which is naturally mirrored in the elongation of synchrotron structures along these curved fields within the 3D spatial space. However, both the observed curvature of the magnetic field lines and the elongation of these structures on the POS are also influenced by the projection effect, providing further insights into the magnetic field’s inclination angle.

\textbf{Magnetization effect:} The degree of observed anisotropy is affected not only by projection effects but also by the medium's magnetization. A higher degree of anisotropy in the intensity structure is more pronounced in environments with strong magnetization, while weaker magnetization not only reduces the anisotropy but also leads to more pronounced curvature in both the magnetic field and the corresponding intensity structure. Thus, understanding the anisotropy degree, along with the topology of the projected intensity structure, offers valuable information on both the magnetization and the inclination angle.

\begin{figure*}
\centering
\includegraphics[width=1.0\linewidth]{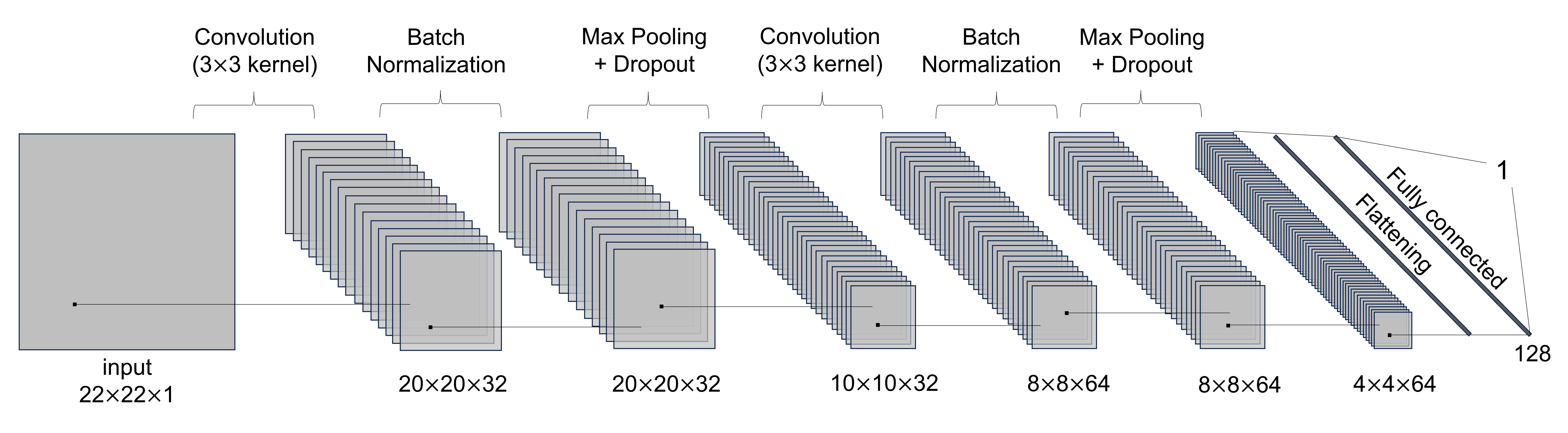}
        \caption{Architecture of the CNN-model. The input image is a $22\times22$-cells map cropped from the synchrotron intensity map. The network outputs the prediction of the magnetic field's POS orientation angle $\phi$, inclination angle $\gamma$, or the Alfv\'en Mach number $M_A$.  Modified from \citet{2024MNRAS.52711240H}.}
    \label{fig:cnn}
\end{figure*}

\section{Numerical method}
\label{sec:data}
\subsection{Convolutional neural network (CNN)}
\subsubsection{CNN architecture}
To construct a deep neural network \citep{lecun1998gradient} to trace the 3D magnetic field from a synchrotron emission map, we adopt a CNN architecture similar to that used in \citet{2024MNRAS.52711240H}. The CNN architecture, as illustrated in Fig.~\ref{fig:cnn}, consists of initial layers comprising a stack of convolutional layers followed by pooling and dropout layers. To facilitate faster convergence during the network training process using backpropagation of the loss and enhance the learning stability, we introduce a batch normalization layer following each convolution layer. After several iterations of convolution and pooling layers, we extract a compressed image feature, which is then processed by the fully connected layers to predict the desired properties. A detailed discussion of each layer's function is given in \citet{2024MNRAS.52711240H}. Such a CNN architecture has been proven to be applicable in tracing 3D magnetic fields using spectroscopic observations. The median uncertainties are under $5^\circ$ for both POS and inclination angles and less than 0.2 for $M_A$ in sub-Alfv\'enic conditions. Compared with \citet{2024MNRAS.52711240H}, our CNN architecture uses fewer convolutional layers, accelerating the training process with less computational time.

\subsubsection{Network Training}
The trainable parameters within the CNN are optimized following a conventional neural network training approach, where the mean-squared error of the 3D magnetic field prediction acts as the training loss for backpropagation. This methodology is grounded in the foundational principles established by \citet{rumelhart1986learning}. 

\textbf{Random cropping:} To bolster the CNN model's ability to generalize, our training strategy includes diversifying the training dataset through data augmentation \citep{doi:10.1198/10618600152418584}. One effective technique is random cropping \citep{2018arXiv181109030T}, which involves generating smaller patches of size $22 \times 22$ cells from the input images. This approach not only expands the dataset but also introduces a variety of perspectives within the data, thereby enhancing the model’s exposure to different features present in the synchrotron emission maps. The size of $22 \times 22$ cells is chosen to avoid numerical dissipation of turbulence and achieve a high-resolution measurement. As shown in Appendix~\ref{app:A}, the size does not affect the CNN's accuracy after sufficient training, but training large patches is more computationally expensive. 

\textbf{Random rotation:} Additionally, images lack rotational invariance from the perspective of the computational model. Each image cell corresponds to an element in a matrix, and rotating an image alters the matrix's element arrangement, presenting the image as novel data to the model \citep{10.1145/1273496.1273556}. This characteristic is exploited in two ways: firstly, by randomly rotating the $22 \times 22$-cell patches to further augment the training dataset, and secondly, by leveraging the original, unrotated datasets for validation, simulating a prediction test scenario.

These augmentation strategies enrich the training dataset with diversity and randomness \citep{doi:10.1198/10618600152418584}, which are crucial for refining CNN's predictive accuracy and generalization across different physical conditions.

\begin{table}
	\centering
 \begin{tabular}{ | c | c | c | c | c | c|}
		\hline
		Run & $M_s$ & $M_A$ & range of $M_A^{\rm sub}$  & range of $M_s^{\rm sub}$ & code\\ \hline \hline
		Z0 & 0.66 & 0.26 & 0.17 - 0.36 & 0.37 - 0.91 &\multirow{5}{*}{ZEUS-MP}\\ 
		Z1 & 0.62 & 0.50 & 0.26 - 0.75 & 0.37 - 0.89 &\\
		Z2 & 0.61 & 0.79 & 0.38 - 1.00 & 0.38 - 0.82 &\\ 
		Z3 & 0.59 & 1.02 & 0.42 - 1.37 & 0.37 - 0.80 &\\ 
		Z4 & 0.58 & 1.21 & 0.49 - 1.55 & 0.38 - 0.82 &\\\hline
  	A0 & 1.21 & 1.25 & 0.51 - 1.56 & 0.58 - 1.53 & AthenaK\\\hline
	\end{tabular}
	\caption{\label{tab:sim}$M_s$ and $M_A$ are the sonic Mach number and the Alfv\'enic Mach number calculated from the global injection velocity, respectively. $M_A^{\rm sub}$ and $M_s^{\rm sub}$ are determined using the local velocity dispersion calculated along each LOS in a $22\times22$ cells sub-field. 
 }
\end{table}

\subsection{MHD simulations}
The MHD numerical simulations presented in this study were generated from the ZEUS-MP/3D and AthenaK code, as detailed by \citet{2006ApJS..165..188H} and \citet{2020ApJS..249....4S}, respectively. We executed an isothermal simulation of MHD turbulence, employing the ideal MHD equations within an Eulerian framework, complemented by periodic boundary conditions. Kinetic energy injection was solenoidally applied at wavenumber 2 to emulate a Kolmogorov-like power spectrum. The turbulence was actively driven until achieving a state of statistical equilibrium. The computational domain was discretized into a $792^3$ cell grid, with numerical dissipation of turbulence occurring at scales between approximately 10 to 20 cells. See \citet{2024MNRAS.527.3945H} for more details.

Initial conditions for the simulations featured a uniform density field and a magnetic field oriented along the $y$-axis. The simulation cubes were subsequently rotated to align the mean magnetic field inclination with respect to the LOS, or the $z$-axis, at angles of $90^\circ$, $60^\circ$, and $30^\circ$, respectively. Characterization of the scale-free turbulence within the simulations was achieved through the sonic Mach number, $M_s = \delta v_{\rm inj}/c_s$, and the Alfv\'enic Mach number, $M_A = \delta v_{\rm inj}/v_A$. To explore various physical scenarios, the initial density and magnetic field settings were adjusted, producing a range of $M_A$ and $M_s$ values. The simulations are referenced throughout this paper by their designated model names or key parameters, as enumerated in Table~\ref{tab:sim}.
\begin{figure*}
\includegraphics[width=1.0\linewidth]{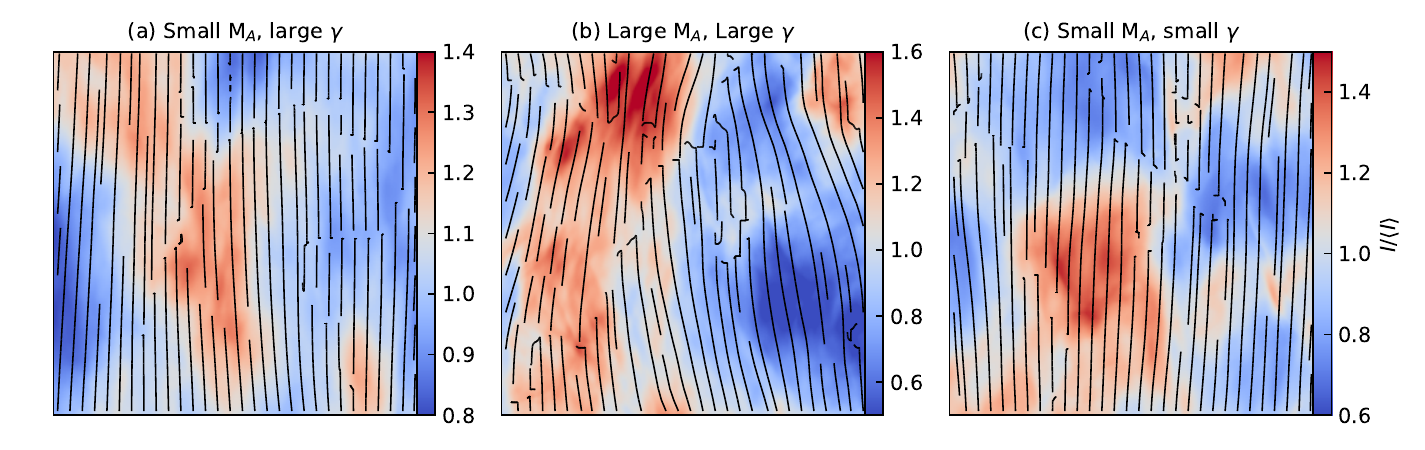}
        \caption{A numerical illustration of the anisotropy in normalized synchrotron intensity map. The black streamlines represent the POS magnetic field orientation. Panel (a): $\langle M_A\rangle=0.26$, $\langle M_s\rangle=0.66$, and $\langle \gamma\rangle=90^\circ$. Panel (b): $\langle M_A\rangle=1.21$, $\langle M_s\rangle=0.58$, and $\langle \gamma\rangle=90^\circ$. Panel (c): $\langle M_A\rangle=0.26$, $\langle M_s\rangle=0.66$, and $\langle \gamma\rangle=30^\circ$. }
    \label{fig:map}
\end{figure*} 
\begin{figure*}
\centering
\includegraphics[width=0.99\linewidth]{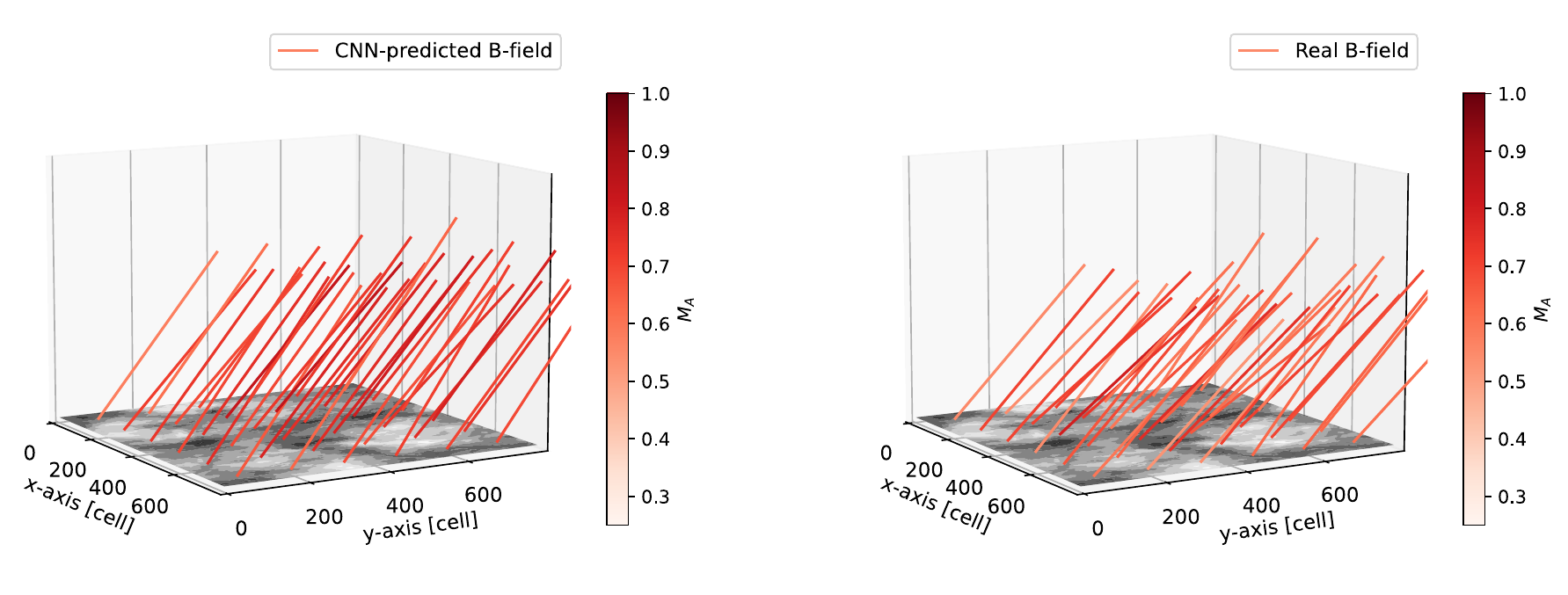}
        \caption{A comparison of the CNN-predicted 3D magnetic fields using the simulation $\langle M_A\rangle=0.79$, $\langle M_s\rangle=0.61$, and $\langle \gamma\rangle=60^\circ$. Each magnetic field segment is constructed by the POS magnetic field's position angle (i.e., $\phi$) and the inclination angle $\gamma$. Note that the magnetic field obtained is the projection along the LOS and averaged over 132$\times$132 pixels for visualization purposes. The third axis of the LOS is only for 3D visualization purposes and does not provide distance information here. The total intensity map $I$ is placed on the POS, i.e., the $x-y$ plane. }
    \label{fig:3D_sub13}
\end{figure*}
\subsection{Synthetic synchrotron observation}
To generate a synthetic synchrotron observation from our simulations, we utilize the density field, $\rho(\mathbf{x})$, and the magnetic field, $\mathbf{B}(\mathbf{x})$, where $\mathbf{x}=(x,y,z)$ denotes the spatial coordinates. The calculation for synchrotron intensity $I(x,y)$ follows \citep{1986rpa..book.....R,1970ranp.book.....P,2016ApJ...831...77L}:
\begin{equation}
\label{eq.iqu}
I(x,y) = \int n_{e}(B_x^2 + B_y^2)B_{\perp}^\alpha dz,
\end{equation}
where $n_{e}=\rho(\mathbf{x})$ is the density of relativistic electrons. Compared with the original definition given in \cite{1986rpa..book.....R,1970ranp.book.....P,2016ApJ...831...77L}, the wavelength-dependent term, as well as other constant factors, are ignored, since it does not change the properties of the fluctuations in a scale-free MHD turbulence simulation.

Considering the anisotropy in synchrotron emission is relatively insensitive to the electron energy distribution's spectral index \citep{2012ApJ...747....5L, 2019ApJ...886...63Z}, we assume a homogeneous and isotropic electron energy distribution $N(E)dE = N_0E^{-p}dE$ with a spectral index $p=3$, where $N_0$ is the pre-factor of the electron distribution. This assumption yields a synchrotron emission index of $\alpha = (p - 3)/4$. The magnetic field predicted from CNN using $I(x,y)$ is insensitive to Faraday rotation. We therefore did not include the Faraday rotation effect here for comparison purposes, while in real observations polarization synchrotron emission is contaminated by Faraday rotation.

\subsection{Training images}
Our training input is the synchrotron intensity map $I(x,y)$ generated from the ZEUS-MP/3D simulations, while the AthenaK simulation serves as a validation test. The intensity map is normalized by its maximum intensity so only morphological features in the map are the most important. The $792\times792$-cells $I(x,y)$ is randomly segmented into $22\times22$-cell subfields for input into the CNN model. For each subfield, we also generate corresponding projected maps of $\phi^{\rm sub}$, $\gamma^{\rm sub}$, $M_A^{\rm sub}$, and $M_s^{\rm sub}$ as per the following:
\begin{equation}
\begin{aligned}
\phi^{\rm sub}(x,y)&=\arctan(\frac{\int B_y(x,y,z)dz}{\int B_x(x,y,z)dz}),\\
\gamma^{\rm sub}(x,y)&=\arccos(\frac{\int B_z(x,y,z)dz}{\int B(x,y,z)dz}),\\
M_A^{\rm sub}&=\frac{v_{\rm inj}^{\rm sub}\sqrt{4\pi\langle\rho\rangle^{\rm sub}}}{\langle B\rangle^{\rm sub}}, \\
M_s^{\rm sub}&=\frac{v_{\rm inj}^{\rm sub}}{c_s},
\end{aligned}
\end{equation}
where $B=\sqrt{B_x^2+B_y^2+B_z^2}$ is the total magnetic field strength, and $B_x$, $B_y$, and $B_z$ are its $x$, $y$, and $z$ components. $\langle\rho\rangle^{\rm sub}$ and $\langle B\rangle^{\rm sub}$ are the mass density and magnetic field strength averaged over the subfield, respectively. $M_A^{\rm sub}$ and $M_s^{\rm sub}$ are defined using the local velocity dispersion for each subfield (i.e., $v_{\rm inj}^{\rm sub}$), rather than the global turbulent injection velocity $v_{\rm inj}$ used to characterize the full simulation. The ranges of $M_A^{\rm sub}$ and $M_s^{\rm sub}$ averaged over the subfield in each simulation with different $\gamma$ are listed in Tab.~\ref{tab:sim}, while $\gamma^{\rm sub}$ spans from 0 to 90$^\circ$. These values of $M_A^{\rm sub}$, $M_s^{\rm sub}$, and $\gamma^{\rm sub}$ cover typical physical conditions of diffuse medium.

\section{Results}
\label{sec:result}
\subsection{Numerical training and tests}
Fig.~\ref{fig:map} shows how the Alfv\'enic Mach number ($M_A$) and the inclination angle ($\gamma$) shape the anisotropy within synchrotron intensity maps, particularly focusing on local intensity structures. When  $M_A$ is small and $\gamma$ is large, representing a strong magnetic field and insignificant projection effect, the intensity structures prominently emerge as narrow strips, aligning with the POS magnetic fields. With an increase in $M_A$, corresponding to a weakening in the magnetic field, both the magnetic field topology and the synchrotron intensity structures exhibit increased curvature.
\begin{figure*}
    \centering
    \subfigure[$\langle M_A\rangle=0.79$, $\langle M_s\rangle=0.61$, $\langle \gamma\rangle=90^\circ$]{
    \centering
    \includegraphics[width=1.0\linewidth]{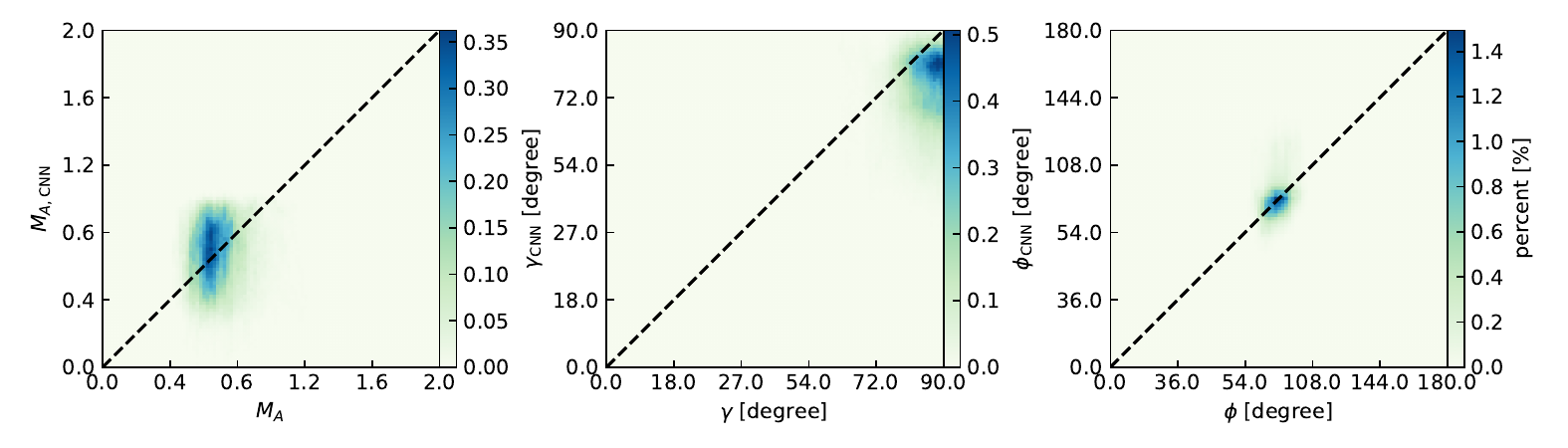}
    }
    \subfigure[$\langle M_A\rangle=0.79$, $\langle M_s\rangle=0.61$, $\langle \gamma\rangle=60^\circ$]{
    \centering
    \includegraphics[width=1.0\linewidth]{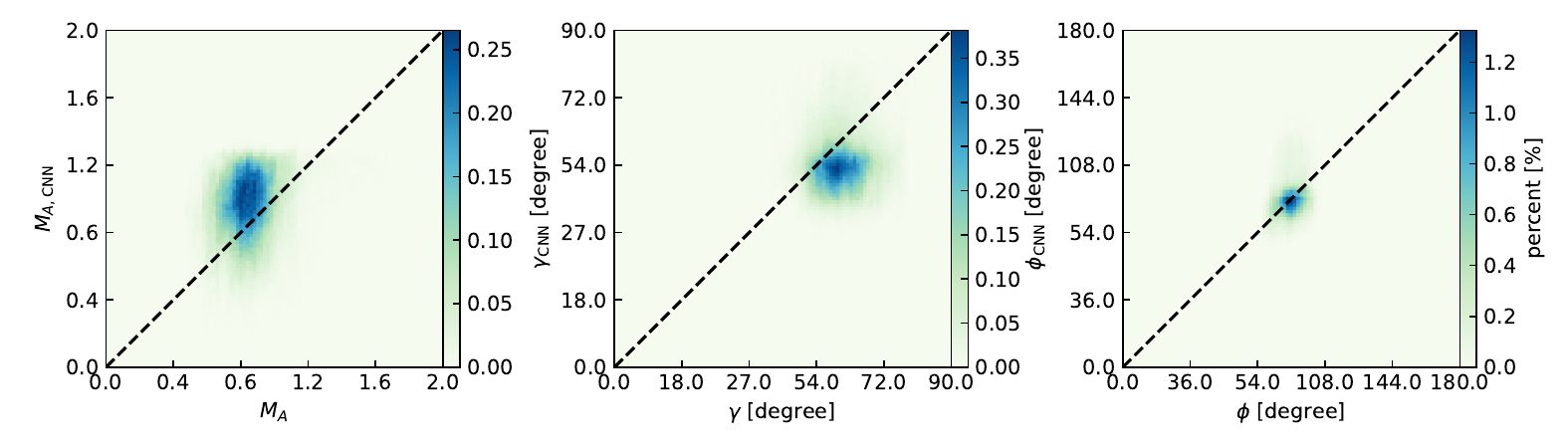}
    }
    \subfigure[$\langle M_A\rangle=1.25$, $\langle M_s\rangle=1.21$, $\langle \gamma\rangle=60^\circ$]{
    \centering
    \includegraphics[width=1.0\linewidth]{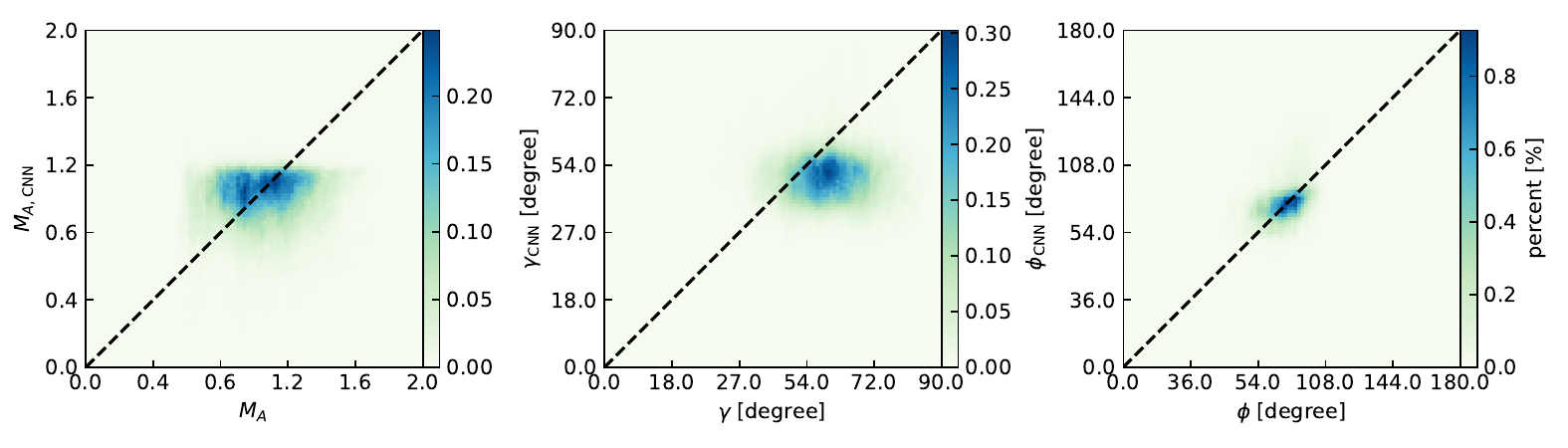}
    }
    \caption{2D histograms of the CNN-predictions, i.e., $\phi^{\rm CNN}$ (right), $\gamma^{\rm CNN}$ (middle), and $M_A^{\rm CNN}$ (left) and the corresponding actual values in simulation. The dashed reference line represents the ideal scenario, where the predicted values and actual values match perfectly.}
    \label{fig:2Dhist}
\end{figure*}

On the other hand, small $\gamma$ suggests a magnetic field orientation closer to the LOS, diminishing the observed anisotropy due to projection effects. Consequently, the elongation along the POS magnetic field becomes less distinct, indicating a reduced anisotropic degree. Thus, the characteristics of anisotropic elongation, curvature, and degree within the intensity structures offer insights into the magnetic fields' POS orientation, inclination angle, and magnetization ($M_A^{-1}$), respectively.

Fig.~\ref{fig:3D_sub13} offers a visual comparison between the actual 3D magnetic fields from a simulation characterized by $\langle M_A\rangle=0.79$, $\langle M_s\rangle=0.61$, and $\langle \gamma\rangle=60^\circ$, and those predicted by a trained CNN model. In this figure, the orientation of the POS magnetic field, represented by the position angle ($\phi$), and $\gamma$, are visualized, with the projected $M_A$ values depicted through color coding. A significant observation from this comparison is the congruence in the orientations of the actual and CNN-predicted 3D magnetic fields. The predicted $M_A$ values, however, are observed to be marginally higher—by approximately 0.1 to 0.2—compared to the actual simulation values.

Fig.~\ref{fig:2Dhist} presents 2D histograms of the CNN predictions—$\phi^{\rm CNN}$, $\gamma^{\rm CNN}$, and $M_A^{\rm CNN}$—against the actual values from two distinct test simulations, Z2 ($\langle M_A\rangle=0.79$, $\langle M_s\rangle=0.61$) and A0 ($\langle M_A\rangle=1.25$, $\langle M_s\rangle=1.21$). It is noteworthy that these simulations were generated using different numerical codes: Z2 by ZEUS-MP/3D \citep{2006ApJS..165..188H} and A0 by the AthenaK code \citep{2020ApJS..249....4S}. Importantly, simulation A0, characterized by a higher Mach number than those included in the CNN training dataset, was not utilized during the training phase. Despite the inherent differences in the numerical simulations and the turbulence conditions they represent, the histograms reveal a statistical concordance between the CNN predictions and the actual simulation values. The proximity of the data points to the one-to-one reference line indicates a strong agreement between predicted and true values, highlighting the CNN model's accuracy, albeit with some scatter that reflects deviations from the actual values.
\begin{figure*}
\centering
\includegraphics[width=1.0\linewidth]{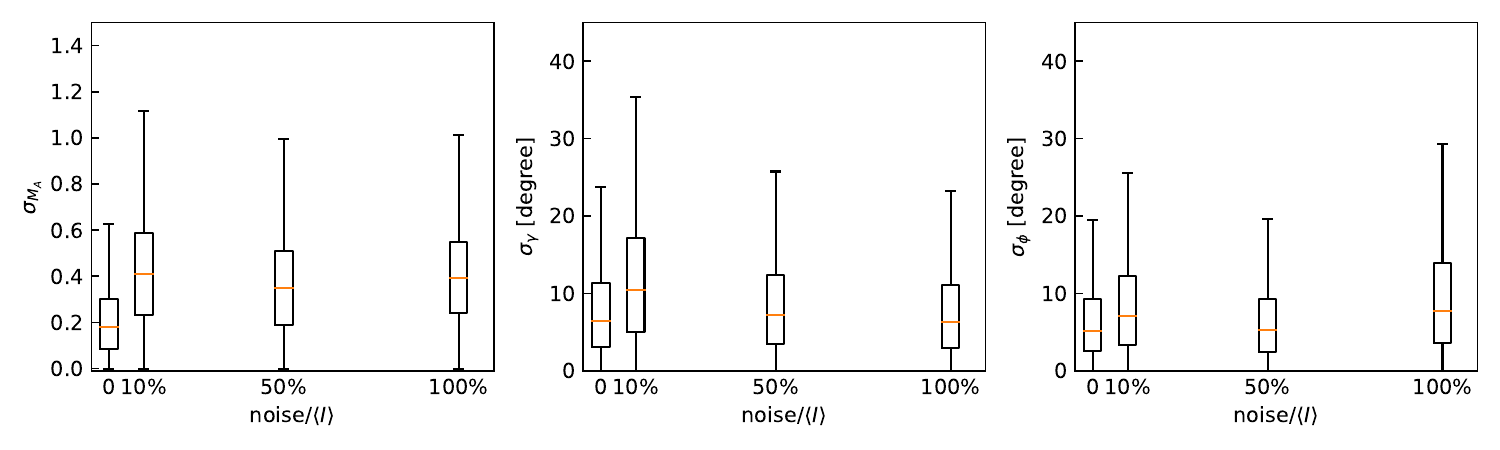}
        \caption{Boxplots of difference in CNN-predicted $\phi^{\rm CNN}$ (right), $\gamma^{\rm CNN}$ (middle), and $M_A^{\rm CNN}$ (left) and the actual values in the simulation A0 ($\langle M_A\rangle=1.25$, $\langle M_s\rangle=1.21$, $\langle \gamma\rangle=60^\circ$) with Gaussian noise introduced. The upper and lower black lines represent the deviation’s maximum and minimum, respectively. The box gives ranges of the first (lower) and third quartiles (upper) and the orange line represents the median value. The amplitude of the noise varies from 10\%, 50\%, to 100\% of the mean intensity of the maps, corresponding to signal-to-noise ratios of 10, 2, and 1, respectively. }
    \label{fig:box_noise}
\end{figure*}

\begin{figure*}
\centering
\includegraphics[width=1.0\linewidth]{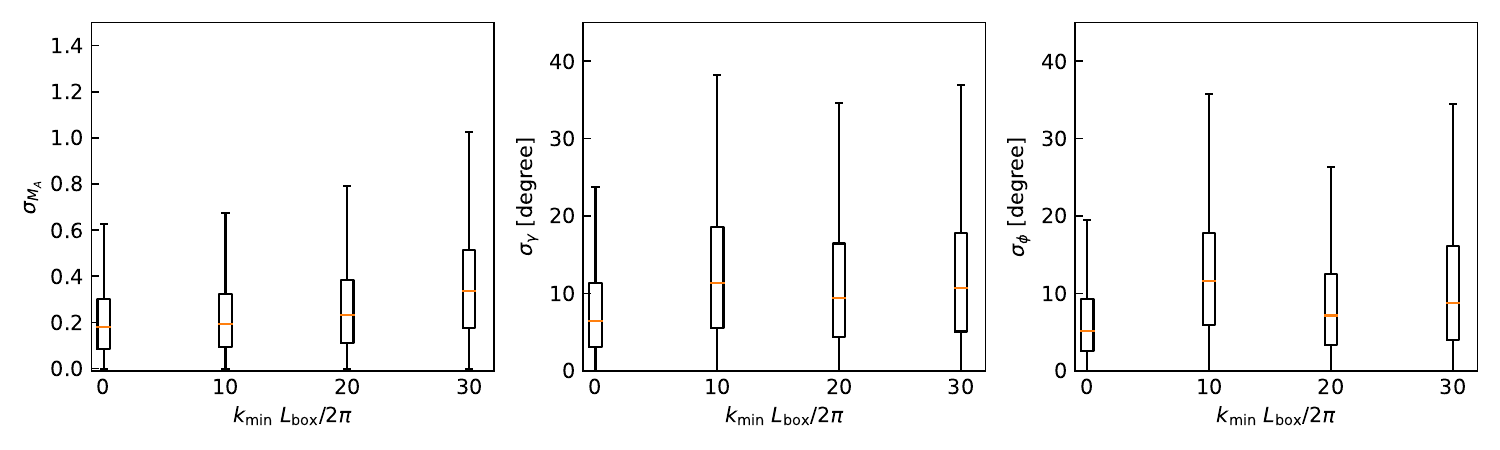}
        \caption{Boxplots of difference in CNN-predicted $\phi^{\rm CNN}$ (right), $\gamma^{\rm CNN}$ (middle), and $M_A^{\rm CNN}$ (left) and the actual values in the simulation A0 ($\langle M_A\rangle=1.25$, $\langle M_s\rangle=1.21$, $\langle \gamma\rangle=60^\circ$) when spat. The upper and lower black lines represent the deviation’s maximum and minimum, respectively. The box gives ranges of the first (lower) and third quartiles (upper) and the orange line represents the median value. $k_{\rm min}$ represents the minimum wavenumber remaining in the filtered synchrotron intensity map and $L_{\rm box}$ is the size of simulation box. }
    \label{fig:box_k}
\end{figure*}

\subsection{Noise effect}
Noise is an inherent challenge in observation that can potentially influence CNN predictions. To evaluate this effect comprehensively, we introduce Gaussian noise into the synchrotron intensity maps used for training the CNN model. The amplitude of this noise varies, representing 10\%, 50\%, and 100\% of the mean intensity of the maps, corresponding to signal-to-noise ratios (SNRs) of 10, 2, and 1, respectively. This allows us to train the CNN model across a range of noise levels.

Fig.~\ref{fig:box_noise} presents boxplots that illustrate the deviations between the CNN-predicted values and the actual 3D magnetic field using simulation A0 ($\langle M_A\rangle=1.25$, $\langle M_s\rangle=1.21$, $\langle \gamma\rangle=60^\circ$) as a case study. We quantify these deviations by calculating the absolute differences in the magnetic field’s position angle ($|\phi^{\rm CNN} - \phi|$), inclination angle ($|\gamma^{\rm CNN} - \gamma|$), and Alfv\'en Mach number ($|M_A^{\rm CNN} - M_A|$), represented as $\sigma_\phi$, $\sigma_\gamma$, and $\sigma_{{\rm M}_A}$, respectively.

In noise-free conditions (see Figs.~\ref{fig:2Dhist} and \ref{fig:box_noise}), the median values of $\sigma_{M_A}$, $\sigma_\gamma$, and $\sigma_\phi$ are approximately 0.2, 5$^\circ$, and 4$^\circ$. Upon introducing noise to the simulation, uncertainties increase, with the median value of $\sigma_{M_A}$ rising to about 0.4, and median $\sigma_\gamma$ and $\sigma_\phi$ extending to the range of 8$^\circ$ - 10$^\circ$. Remarkably, these uncertainties remain consistent across different SNRs of 10, 2, and 1, underscoring the CNN model's ability to extract magnetic field information amidst varying levels of noise.
\begin{figure*}
\centering
\includegraphics[width=1.0\linewidth]{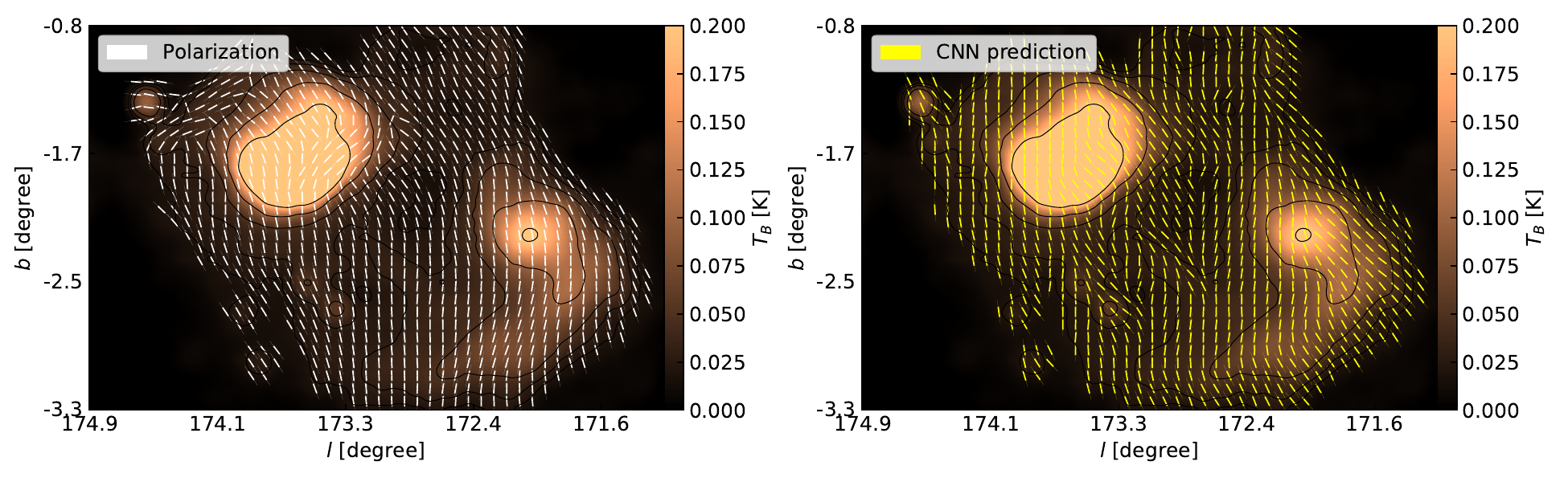}
        \caption{Comparison of the POS magnetic fields predicted by the CNN (right) using the Sino-German $\lambda$6 cm synchrotron emission data and inferred from synchrotron polarization (left). The background image is the integrated synchrotron intensity map. }
    \label{fig:hii_prediction}
\end{figure*}

\begin{figure*}
\centering
\includegraphics[width=1.0\linewidth]{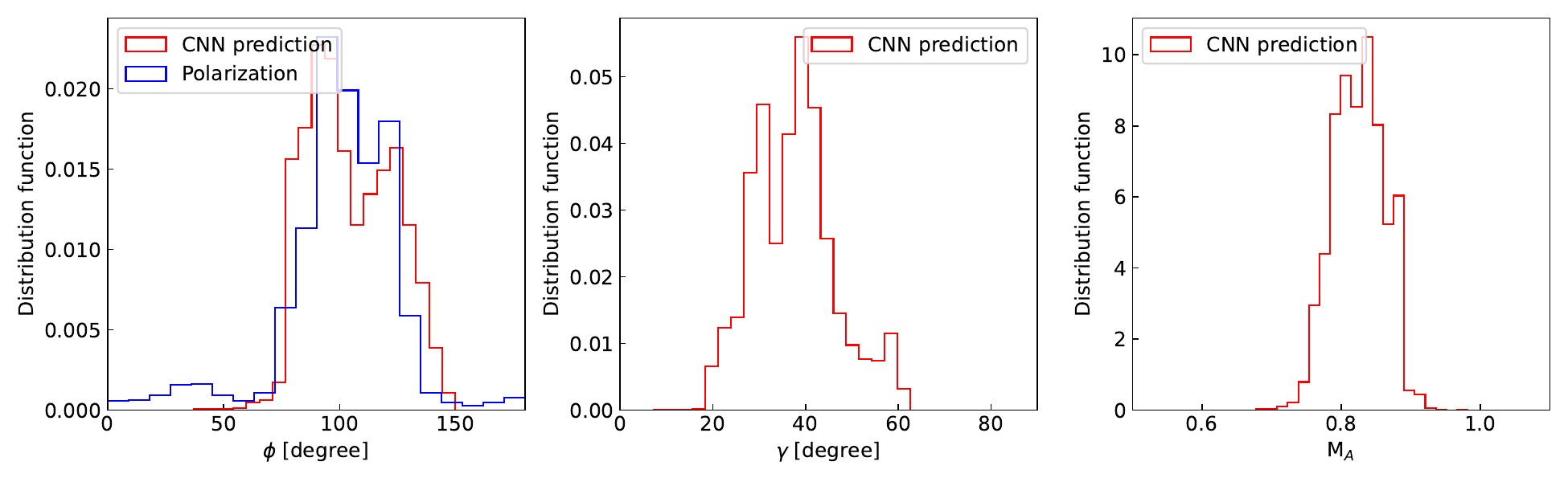}
        \caption{Histograms of CNN-predicted $\phi$ (left), defined north from the west, $\gamma$ (middle), and M$_A$ (right). The synchrotron polarization measured $\phi$ is also given for comparison.}
    \label{fig:hii_hist}
\end{figure*}

\subsection{Removing low-spatial-frequency components}
Traditional magnetic field mapping via polarimetry necessitates a comprehensive range of spatial frequencies, incorporating both the high-spatial-frequency data from interferometers and the low-spatial-frequency data from single-dish observations. However, recent studies \citep{2020arXiv200207996L, 2022arXiv220806074H} have illuminated that the anisotropy inherent in MHD turbulence and synchrotron emission is more pronounced at higher spatial frequencies. This suggests that the CNN approach could effectively obtain the POS orientation, inclination angle, and magnetization using exclusively high-spatial-frequency data. To examine this proposition, we applied a $k$-space filter to synchrotron intensity maps before CNN training, involving: (i) performing a Fast Fourier Transform (FFT) on a 2D map; (ii) filtering out the intensity values at specified wavenumbers $k$ to highlight high-spatial frequencies; and (iii) applying an inverse FFT to transform the filtered map back to the spatial domain.

Fig.~\ref{fig:box_noise} presents boxplots showing the deviations—$\sigma_{M_A}$, $\sigma_\gamma$, and $\sigma_\phi$—resulting from the removal of low-spatial-frequency components in the synchrotron intensity maps. Across different scenarios of wavenumber removal ($k<10$, $k<20$, and $k<30$), the median values of $\sigma_{M_A}$, $\sigma_\gamma$, and $\sigma_\phi$ exhibit some variability but generally maintain stability within $<0.4$, $<10^\circ$, and $<10^\circ$, respectively. This underscores the CNN method's robustness in probing 3D magnetic fields, even without the contribution of low spatial frequencies—highlighting its particular suitability for processing interferometric data devoid of single-dish measurements.

\subsection{Observational application}
For our observational tests, we selected a diffuse emission region from the Sino-German $\lambda$6 cm polarization survey \citep{2010A&A...515A..64G}. This region, situated away from the Galactic plane, minimizes the complexity arising from the LOS projection of multiple components. It also avoids energetic sources such as supernova remnants, which could significantly alter the properties of the surrounding gas. The synchrotron intensity data has a beam resolution of 9.5$'$ and an RMS noise level of 0.5-0.7 mK in brightness temperature $T_B$, while the noise level for polarized intensity is 0.3 - 0.5 mK in $T_B$. 

As shown in Fig.~\ref{fig:hii_prediction}, there is consistency between the POS magnetic field orientations predicted by the CNN and those derived from synchrotron polarization, though misalignment is evident in the northeast clump. This clump is associated with the H II complex SH-236, where radiation and stellar winds from young stars may compress the surrounding gas and alter its physical properties. Currently, our CNN model does not account for the impacts of stellar winds and radiation, which could explain the inaccurate predictions observed in this region. Additionally, Faraday rotation is not corrected for the synchrotron polarization data, which might introduce further uncertainties. 

A significant advantage of our CNN approach over traditional polarization methods is its capability to probe the inclination angle $\gamma$ and magnetization $M_A$. The CNN-predicted $\gamma$ and $M_A$ maps are shown in Fig.~\ref{fig:HII_ma_gamma}. These predictions are summarised in histograms within Fig.~\ref{fig:hii_hist}. According to the histograms, the median $\gamma$ and $M_A$ are estimated at $\approx40^\circ$ and $\approx0.82$, respectively. According to our study of the uncertainties (see Fig.~\ref{fig:box_noise}), there could be $5^\circ-10^\circ$ uncertainty in the predicted median $\gamma$ and 0.2 - 0.4 uncertainty in the predicted median $M_A$.

\section{Discussion}
\label{sec:dis}
\subsection{Comparison with earlier studies}

Exploring 3D magnetic fields within the ISM using CNNs is rapidly progressing. An initiative by \cite{2024MNRAS.52711240H} showcased the use of a CNN model to trace the 3D magnetic field structure in molecular clouds. This achievement was facilitated by CNN's application to thin velocity channel maps \citep{LP00,2023MNRAS.524.2994H} derived from spectroscopic data.

Building on this CNN approach, our study extends the application of CNN to synchrotron emission, aiming to trace the 3D magnetic field in the warm/hot gas phase. This includes determining the orientation of the POS's magnetic field, the field's inclination angle, and the total Alfv\'en Mach number. As the synchrotron intensity is not subject to Faraday rotation, it does not require multiple frequency measurements to compensate for the Faraday effect. Potential applications of the CNN approach extend across a diverse range of astrophysical environments. These include studying the warm ionized phase of the ISM, the Central Molecular Zone (CMZ), external galaxies, supernova remnants, and galaxy clusters. It can provide important information to address fundamental questions related to the origins of ultra-high energy cosmic rays \citep{2014CRPhy..15..339F,2019JCAP...05..004F}, as well as issues concerning Galactic foreground polarization \citep{2015PhRvD..91h1303K,2016A&A...594A..25P}.

\subsection{Application to interferometric observations}
The challenge of missing low-spatial frequency data in observations made with interferometers, due to constraints imposed by their baseline, is a notable concern in radio astronomy. Instances such as the observations by the Australia Telescope Compact Array (ATCA) at 1.4 GHz \citep{2011Natur.478..214G}, which lacked single-dish measurements, and the Westerbork Synthesis Radio Telescope (WSRT) observations of the 3C 196 field at 350 MHz \citep{2015A&A...583A.137J}, where single-dish measurements at the same frequency were unfeasible, underscore this issue. Additionally, data from the Low-Frequency Array (LOFAR) also experience the loss of low-spatial frequencies \citep{2014A&A...568A.101J}.

Notwithstanding these challenges, the absence of low-spatial frequency information does not impede the application of CNNs for probing 3D magnetic fields. This resilience stems from the CNN approach's foundation on the anisotropy inherent in MHD turbulence and synchrotron radiation, which is more conspicuous at higher spatial frequencies \citep{2020arXiv200207996L,2022arXiv220806074H}. As demonstrated in this study, the CNN model is adept at capturing this anisotropy, leveraging only the high-spatial-frequency data accessible from interferometric observations. This unique capability signifies a significant advantage of the CNN approach, enabling the reconstruction of 3D magnetic fields even in the absence of comprehensive spatial frequency coverage.

\subsection{Synergy with other methods}
The CNN approach has been applied to spectroscopic observations to trace the 3D magnetic fields \citep{2024MNRAS.52711240H}. Extending CNNs to both spectroscopic and synchrotron emission data enables an in-depth analysis of the distribution and variation of 3D magnetic fields across different ISM phases. Compared to synchrotron emission, training CNNs for conditions in molecular clouds presents additional complexities due to the influence of self-gravity and outflow feedback on fluid dynamics and magnetic field structures \citep{2012ApJ...761..156F,2019FrASS...6....3H,2022MNRAS.513.2100H,2022ApJ...941...92H,2024arXiv240300744V}. This necessitates the use of nuanced numerical simulations of molecular clouds for CNN training.

The Synchrotron Intensity Gradient (SIG) technique \citep{2017ApJ...842...30L,2020ApJ...901..162H,2024NatCo..15.1006H} offers a parallel approach to tracing the POS magnetic field orientation, rooted in the anisotropy of MHD turbulence evident in synchrotron emission. This anisotropy manifests in both sub-Alfv\'enic and super-Alfv\'enic turbulence, the latter resulting from the advection of turbulent flows \citep{2024NatCo..15.1006H}. With several numerical and observational validation \citep{2017ApJ...842...30L,2019MNRAS.486.4813Z,2020ApJ...901..162H,2024NatCo..15.1006H}, SIG serves as a valuable benchmark for assessing the efficacy of CNN-based models, particularly in scenarios where polarization data are scarce, such as radio halos in galaxy clusters.

Furthermore, it should be noted that the inclination angle predicted by the CNN model is inherently limited to the range of [0, 90$^\circ$]. This limitation arises because the anisotropy alone cannot definitively discern whether the magnetic field is oriented towards or away from the observer. A synergy with Faraday rotation measurements \citep{2007ASPC..365..242H,2009ApJ...702.1230T,2012A&A...542A..93O,2016ApJ...824..113X,2019A&A...632A..68T}, offers promising avenues to resolve this degeneracy. On the other hand, the proposed CNN method is based on the dominance of MHD turbulence in synchrotron emission statistics. When other physical processes are important, including inflow and outflow, additional training data sets are required.

\section{Summary}
\label{sec:con}

In this study, we developed and evaluated a CNN model designed to investigate three-dimensional (3D) magnetic fields, including the orientation of the POS magnetic field, the field's inclination angle, and total magnetization, utilizing synchrotron intensity maps. Our major findings are summarized as follows:
\begin{enumerate}
    \item We designed and implemented a CNN model capable of extracting the orientation of the POS magnetic field, the field's inclination angle, and the overall magnetization from synchrotron intensity maps.
   \item Through the utilization of synthetic synchrotron maps for model training, we identified that the median uncertainties for predicting the magnetic field’s position angle ($\phi$) and inclination angle ($\gamma$) remained below $10^\circ$, with the Alfv\'en Mach number ($M_A$) uncertainty staying under $0.4$. 
    \item The model's robustness against noise was evaluated, demonstrating insensitivity to noise with adequate training, ensuring reliable performance under various observational conditions.
    \item Our analyses confirmed the CNN method's applicability in tracing the POS magnetic field orientation, the field's inclination angle, and total magnetization, even in the absence of low spatial frequencies in the synchrotron images—making it particularly adept for analyzing interferometric data that lacks single-dish measurements.
    \item We tested this trained CNN model by applying it to the synchrotron observations of a diffuse region. The CNN-predicted POS magnetic field orientation shows a statistical agreement with that derived from synchrotron polarization.
    \item We discussed the potential and future applications of this CNN method, particularly its utility in predicting the 3D Galactic Magnetic Fields (GMF), and its implications for comprehending 3D magnetic fields within the Central Molecular Zone (CMZ) and beyond, in external galaxies.
\end{enumerate}

\begin{acknowledgments}
Y.H. acknowledges the support for this work provided by NASA through the NASA Hubble Fellowship grant No. HST-HF2-51557.001 awarded by the Space Telescope Science Institute, which is operated by the Association of Universities for Research in Astronomy, Incorporated, under NASA contract NAS5-26555. A.L. acknowledges the support of NASA ATP AAH7546, NSF grants AST 2307840, and ALMA SOSPADA-016. Financial support for this work was provided by NASA through award 09\_0231 issued by the Universities Space Research Association, Inc. (USRA). This work used SDSC Expanse CPU at SDSC through allocations PHY230032, PHY230033, PHY230091, and PHY230105 from the Advanced Cyberinfrastructure Coordination Ecosystem: Services \& Support (ACCESS) program, which is supported by National Science Foundation grants \#2138259, \#2138286, \#2138307, \#2137603, and \#2138296. We acknowledge ChatGPT's contribution in proofreading the manuscript.
\end{acknowledgments}

%


\vspace{15mm}
\software{AthenaK code \citep{2020ApJS..249....4S}; ZEUS-MP/3D code \citep{2006ApJS..165..188H}; Python3 \citep{10.5555/1593511}; TensorFlow \citep{tensorflow2015-whitepaper}}


\appendix
\section{Comparison of CNN's input patch size}
\label{app:A}
\begin{figure}
\centering
\includegraphics[width=0.5\linewidth]{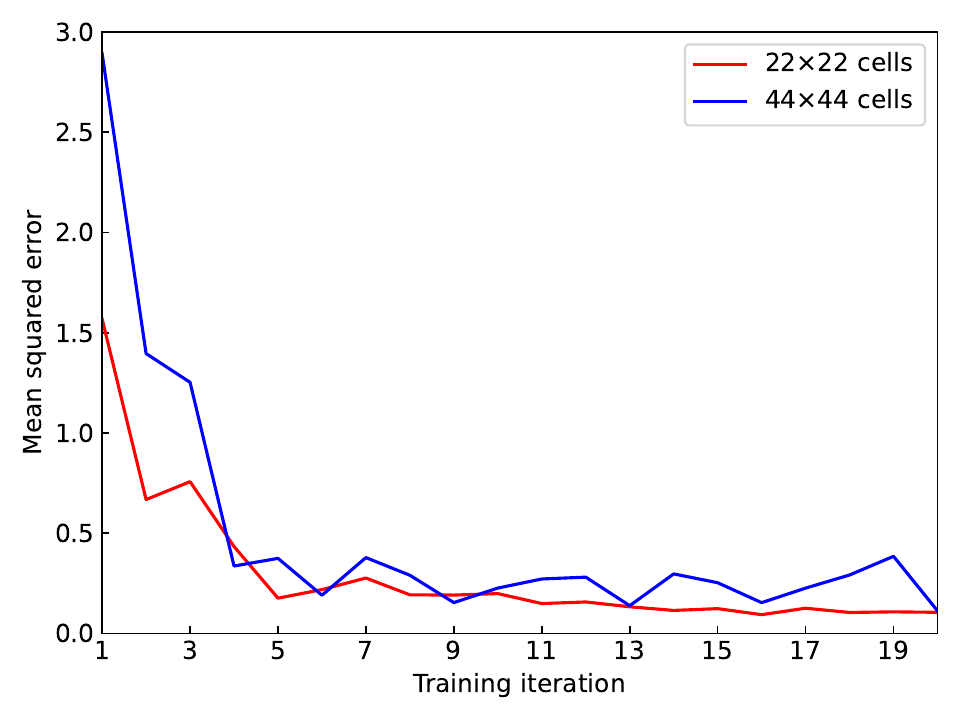}
        \caption{Evolution of CNN's validation loss, i.e., the mean squared error. Two cases with an input patch size of $22\times22$ cells or $44\times44$ cells are tested. }
    \label{fig:loss}
\end{figure}
Fig.~\ref{fig:loss} shows the variation in validation loss for two different input patch sizes, $22\times22$ cells and $44\times44$ cells. The validation loss, representative of the mean squared error between the predicted and actual 3D magnetic fields, is derived from validation datasets comprising patches randomly extracted from the Zeus-series simulations (see Tab.~\ref{tab:sim}). For each training iteration, 100,000 patches are utilized to compute the validation loss, with the loss being averaged across the magnetic field's POS angle, inclination angle, and Alfv\'en Mach number.
We can see regardless of the input patch size, whether $22\times22$ or $44\times44$ cells, the validation loss exhibits a comparable downward trend towards a similar level after an adequate number of training iterations.

\section{Maps of CNN-predicted inclination angle and magnetization}
\label{app:B}
\begin{figure*}
\includegraphics[width=1.0\linewidth]{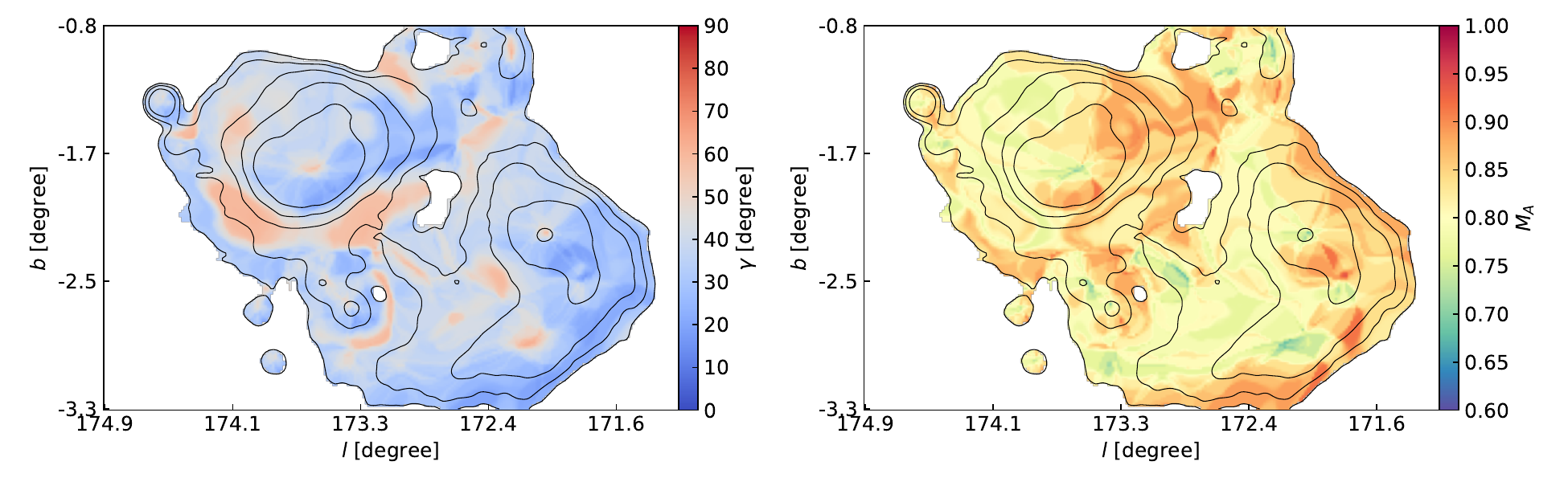}
        \caption{Maps of CNN-predicted inclination angle $\gamma$ (left panel) and Alfv\'en Mach number $M_A$ using the Sino-German $\lambda$6 cm synchrotron emission data. The contours represent the synchrotron intensity structures seen in Fig.~\ref{fig:hii_prediction}.}
    \label{fig:HII_ma_gamma}
\end{figure*}

The maps of CNN-predicted inclination angle $\gamma$ and Alfv\'en Mach number $M_A$ are shown in Fig.~\ref{fig:HII_ma_gamma}. The Sino-German $\lambda$6 cm synchrotron emission data is used. According to the histograms shown in Fig.~\ref{fig:hii_hist}, the median $\gamma$ and $M_A$ are estimated at $\approx40^\circ$ and $\approx0.82$, respectively.


\bibliography{sample631}{}
\bibliographystyle{aasjournal}



\end{document}